# Single-bubble sonoluminescence: Shape stability analysis of collapse dynamics in a semianalytical approach

Vladislav A. Bogoyavlenskiy*

*Low Temperature Physics Department, Moscow State University, 119899 Moscow, Russia*
(Received 29 December 1999; revised manuscript received 21 April 2000)

This paper theoretically analyzes the hydrodynamic shape stability problem for sonoluminescing bubbles. We present a semianalytical approach to describe the evolution of shape perturbations in the strongly nonlinear regime of violent collapse. The proposed approximation estimating the damping rate produced by liquid viscosity is used to elucidate the influence of the collapse phase on subsequent evolution of the Rayleigh-Taylor instability. We demonstrate that time derivatives of shape perturbations grow significantly as the bubble radius vanishes, forming the dominant contribution to destabilization during the ensuing bounce phase. By this effect the Rayleigh-Taylor instability can be enhanced drastically, yielding a viable explanation of the upper threshold of driving pressure experimentally observed by Barber *et al.* [Phys. Rev. Lett. **72**, 1380 (1994)].

## I. INTRODUCTION

In the last decade, a significant impulse to theoretical and experimental studies of the bubble collapse problem was given by the discovery of the single-bubble sonoluminescence phenomenon [1–4]. If a gas bubble in water is subjected to a periodic spherical sound wave of ultrasonic frequency, the acoustic energy can be concentrated by over 12 orders of magnitude in very small volume. During the rarefaction part of the acoustic cycle the bubble absorbs energy from the sound wave, and the subsequent compressional portion of the sound field causes the collapse; the resulting excitation and heating of the gas inside the bubble may lead to UV-light emission of picosecond duration. One of the remarkable features of sonoluminescing bubbles observed by Putterman and co-workers is high sensitivity of the light emission to experimental conditions such as forcing pressure, ambient bubble radius, water temperature, and type of gas mixture [4–8]. Optical measurements reveal different dynamic regimes of bubble behavior, and stable sonoluminescence is found only in a narrow range of external parameters. Particularly puzzling is the dependence on the amplitude of the forcing pressure where an upper threshold effect was reported [5]. Usually, the emission of light takes place when the amplitude of the sound wave exceeds the edge of sonoluminescence; if the sound intensity is increased further, beyond a threshold, the light is quenched.

In order to describe the nontrivial experimental results, Brenner *et al.* introduced the concept that the observed upper threshold marks the onset of shape instabilities on the bubble surface [9,10]. On the basis of linear hydrodynamic analysis, they argued that the strongest destabilization develops when the bubble radius reaches its minimum. The acceleration of compressed gas into the surrounding liquid is enormous, motivating the Rayleigh-Taylor instability that causes exponential growth of shape perturbations on time scales of less than $10^{-9}$ s [10]. This theory, supported by theoretical [11–13] and experimental [14,15] investigations, was nethertheless criticized by Putterman and co-workers as to its background; they claimed that under experimental conditions the liquid viscosity would quench the shape perturbations, so it is necessary to find some mechanism other than the Rayleigh-Taylor instability that results in the quenching of sonoluminescence [7,8] (different points of view on the problem are published in Refs. [16,17]). The posed discrepancies were recently examined by numerical simulations of the full hydrodynamic model considering the viscous nonlocal effects [18,19]. Although these studies have demonstrated a satisfactory agreement between the exact hydrodynamics and its approximation [9,10], further theoretical clarifications still seem desirable.

In the present work, we propose a semianalytical approach to clarify the shape stability problem for sonoluminescing bubbles. Our goal is formulated as a detailed investigation of shape perturbations in the region of the violent collapse preceding the intensive development of the Rayleigh-Taylor instability. For this purpose, analytical solution of the Rayleigh-Plesset equation modeling the liquid viscosity [20] is used to derive the perturbation dynamics as a single relation (distortion amplitude vs bubble radius) appropriate for subsequent theoretical analysis. We demonstrate that time derivatives of the shape perturbations can grow drastically as the bubble collapses, giving the dominant contribution to posterior evolution of the Rayleigh-Taylor instability during the shocklike bounce. This allows us to elucidate the destabilization mechanism leading to the upper threshold effect [5], and also to estimate the influence of liquid viscosity on the shape stability. The paper is organized as follows. In Sec. II, we propose and justify the analytical approximation of the bubble dynamics for the violent collapse phase. The subject of Sec. III is the derivation of the perturbation dynamics for collapsing bubbles. In Sec. IV, we analyze evolution of the shape perturbations during the collapse, and then give a phenomenological description of the Rayleigh-Taylor instability development leading to the quenching of sonoluminescence. Finally, Sec. V formulates a summary of the results obtained.

*Electronic address: bogoyavlenskiy@usa.net

## II. BUBBLE DYNAMICS

### A. The Rayleigh-Plesset equation

Since Lord Rayleigh treated the collapse of an empty cavity in an inviscid liquid [21], much refinement has been done in the theory of bubble dynamics [22]. The main step was the introduction of liquid viscosity, surface tension, and variable external driving pressure by Plesset [23]. Following this formulation, the motion of the bubble wall $R(t)$ obeys the relation (named the Rayleigh-Plesset equation)

$$\rho\left(\ddot{R}R + \frac{3}{2}\dot{R}^2\right) + 4\mu\frac{\dot{R}}{R} + \frac{2\sigma}{R} + \left(1 + \frac{R}{c}\frac{d}{dt}\right)(P_0 + P_a - P_g)$$
$$= 0. \tag{1}$$

Here overdots denote time derivatives; $\rho, \mu,$ and $\sigma$ are the density, shear viscosity, and surface tension coefficient of the liquid, respectively; $c$ is the sound speed in the liquid; $P_0$ = const is the ambient hydrostatic pressure; $P_a$ is the driving acoustic pressure; and $P_g$ is the gas pressure inside the bubble. For sonoluminescing bubbles, the external sound field $P_a$ represents a spatially homogeneous, standing wave:

$$P_a = -P_a^0 \sin 2\pi\omega t, \tag{2}$$

where $P_a^0$ is the amplitude and $\omega$ is the frequency of the acoustic field.

A key aspect of modeling the Rayleigh-Plesset dynamics is the specification of the internal pressure $P_g$. The problem consists in the complexity of the thermofluid mechanical processes such as heat transport at the bubble-liquid interface and formation of shock waves inside the bubble [24–29]. It should be mentioned, however, that in an early paper Trilling [24] concluded that these shock waves would not significantly affect the pressure variation at the bubble wall, which is the primary determinant of the radial motion; as also demonstrated by Prosperetti et al. [25,26], at moderate pressure amplitudes the temperature variations of the liquid near the bubble are negligible. As a consequence, at conditions relevant to sonoluminescing bubbles the gas pressure $P_g$ can be considered to obey the van der Waals process equation, giving a rather precise resemblance between theoretical curves $R(t)$ and experimentally obtained data [30]:

$$P_g = P_0\left(\frac{R_0^3 - h^3}{R^3 - h^3}\right)^k. \tag{3}$$

Here $R_0$ is the ambient bubble radius, $h$ is the collective hard core van der Waals radius, and $k$ is the effective polytropic exponent varying from 1 (the isothermal condition) to the ratio of specific heats $\gamma$ (the adiabatic condition).

In Fig. 1, we present a typical example of the dynamics $R(t)$ simulated for an air bubble in water; the values of parameters in Eqs. (1)–(3) are chosen to satisfy an experimental regime where sonoluminescence is observed [31]. For one acoustic period $T = 37.7$ $\mu$s, one can resolve three distinct stages of the bubble dynamics. During the first part of the acoustic cycle, $0 < t(\mu s) \lesssim 17.5$, the bubble radius expands from its ambient value $R_0 = 4.5$ $\mu$m, to the maximum $R_{max} = 47.1$ $\mu$m. Then the collapse phase, $17.5 < t(\mu s) \lesssim 21.8$, takes place; at the end of the compression, a sharp peak of UV light is emitted as the bubble radius approaches the minimum $R_{min} = 0.56$ $\mu$m. After the collapse, there is the stage of weak secondary oscillations, $21.8 < t(\mu s) \lesssim T$; during this phase, the bubble dissipates the energy accumulated from the sound field by viscous damping, and its radius approaches the ambient value $R_0$ by the beginning of the next acoustic cycle.

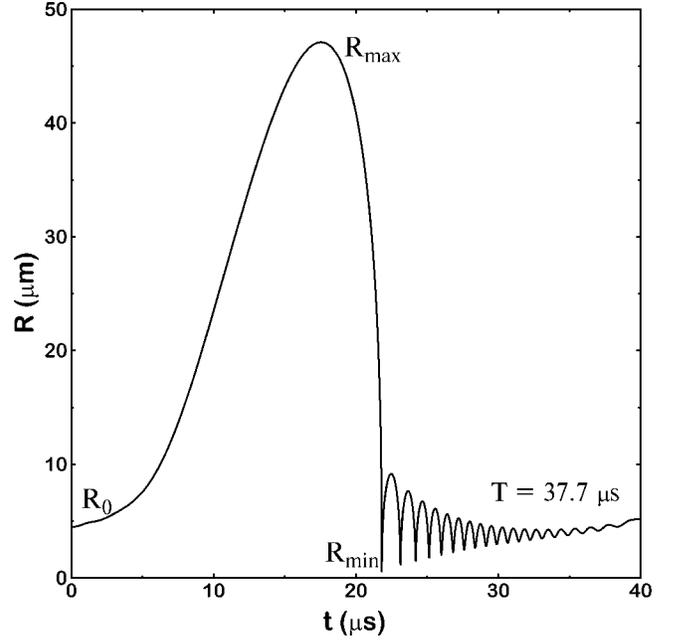

FIG. 1. Dynamics $R$ vs $t$ for an air bubble in water during one acoustic period $T$. Numerical simulation of the Rayleigh-Plesset equation corresponds to an experimental regime where sonoluminescence is observed: $R_0 = 4.5$ $\mu$m, $P_0 = 1$ atm, $P_a^0 = 1.325$ atm, and $\omega = 26.5$ kHz. The material constants are $\rho = 1$ g/cm$^3$, $\mu = 0.01$ g/cm s, $\sigma = 73$ g/s$^2$, $c = 1481$ m/s, $R_0/h = 8.5$, and $k = 1.4$.

Among the three stages of the bubble dynamics, we focus on the collapse phase responsible for significant accumulation of sound energy. In order to discuss this region in detail, we rewrite Eq. (1) as

$$-\rho\ddot{R}R = (P_{vel} + P_{ext} + P_{sur}) - (P_{vis} + P_{gas}), \tag{4}$$

where

$$P_{vel} \equiv \frac{3\rho\dot{R}^2}{2}, P_{ext} \equiv P_0 + P_a + \frac{R\dot{P}_a}{c}, P_{sur} \equiv \frac{2\sigma}{R}, \tag{5}$$

$$P_{vis} \equiv -\frac{4\mu\dot{R}}{R}, P_{gas} \equiv P_g + \frac{R\dot{P}_g}{c}. \tag{6}$$

In this representation, we separate the terms in Eq. (1) that either accelerate ($P_{vel}, P_{ext}, P_{sur}$) or decelerate ($P_{vis}, P_{gas}$) the bubble wall motion. The overall picture demonstrating the contributions of the accelerating and decelerating terms for the discussed Rayleigh-Plesset dynamics (Fig. 1) is summarized by Fig. 2: (a) shows the dependence $R(t)$ during the collapse phase; in plots (b) and (c), we present the evolution of pressures $P_{vel}, P_{ext}, P_{sur}, P_{vis},$ and $P_{gas}$. As illustrated by the auxiliary lines, the collapse region can be conditionally subdivided into three intervals: (i) the weak collapse, (ii) the violent collapse, and (iii) the bounce; within these intervals, the $P_{ext}$, $P_{vel}$, and $P_{gas}$ pressure terms, respectively, are dominant in Eq. (4).

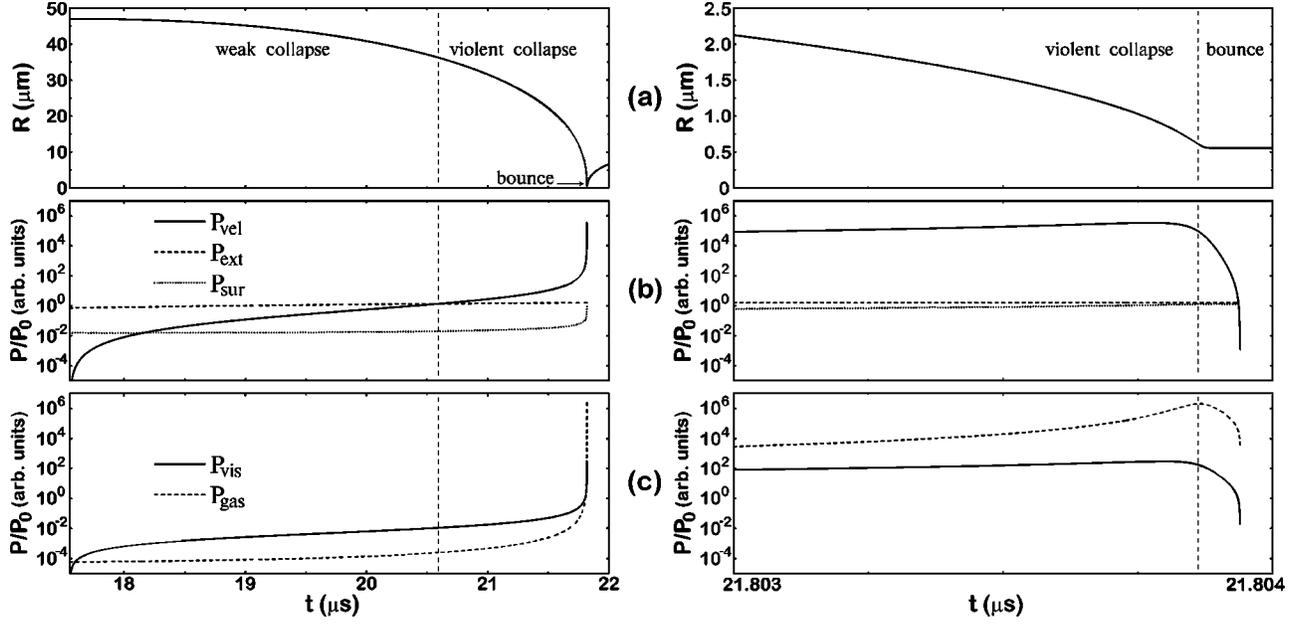

FIG. 2. Collapse phase [$17.5 < t(\mu s) \lesssim 21.8$] of the Rayleigh-Plesset equation simulated with the same parameters as in Fig. 1; (a) bubble dynamics $R$ vs $t$; (b) $\log_{10} P/P_0$ vs $t$ for accelerating pressures $P_{vel}$, $P_{ext}$, and $P_{sur}$; (c) $\log_{10} P/P_0$ vs $t$ for decelerating pressures $P_{vis}$ and $P_{gas}$. Left plots present the regions of weak and violent collapses; right plots, the bounce region in expanded time scale.

### B. The violent collapse phase

Since our goal is the investigation of the bubble shape stability during the violent collapse phase, we need to formulate an adequate approximation of the dynamics $R(t)$ in this region. From Fig. 2, the term $P_{vel}$ gives the dominant contribution to the bubble wall acceleration so the dynamics $R(t)$ principally follows from the classic Rayleigh equation [21]:

$$\ddot{R}R + \frac{3}{2}\dot{R}^2 = 0. \tag{7}$$

To derive the next approximation, we take into consideration the viscosity term $P_{vis}$, which dominates among the decelerating pressures until the shocklike bounce emerges [Fig. 2(c)]. As a result, we introduce the following simplification of the Rayleigh-Plesset equation (see Appendix A):

$$\ddot{R}R + \frac{3}{2}\dot{R}^2 + \frac{4\mu}{\rho}\frac{\dot{R}}{R} = 0, \tag{8}$$

with initial conditions

$$R\{t = t_i\} = R_i, \dot{R}\{t = t_i\} = -V_i. \tag{9}$$

Here $t_i$, $R_i$, and $V_i$ are the initial time, bubble radius, and bubble wall velocity, respectively, related to the beginning of the violent collapse ($P_{vel} \sim P_{ext}$); to improve the fit, the velocity $V_i$ should slightly exceed its actual value $-dR/dt\{t = t_i\}$, as discussed in Appendix A.

The introduced approximation of the Rayleigh-Plesset dynamics [Eqs. (8) and (9)] is integrable, giving the analytical dependence between the bubble radius and time [20]:

$$\frac{4\mu}{\rho R_i^2}(t - t_i) = \frac{1}{4}(1 - \tilde{R}^2) - \frac{\alpha}{3}(1 - \tilde{R}^{3/2}) + \frac{\alpha^2}{2}(1 - \tilde{R})$$

$$- \alpha^3(1 - \tilde{R}^{1/2}) - \alpha^4 \ln\frac{\alpha + \tilde{R}^{1/2}}{\alpha + 1}, \tag{10}$$

where $\tilde{R} \equiv R/R_i$ is the dimensionless bubble radius and $\alpha$ is the parameter of liquid viscosity defined by the relation

$$\alpha \equiv \frac{\rho R_i V_i}{8\mu} - 1. \tag{11}$$

In the case of $\alpha > 0$, Eq. (10) describes the dynamics of viscous collapse,

$$\dot{R} = -\left(\frac{8\mu}{\rho}\right)\frac{1 + \alpha \tilde{R}^{-1/2}}{R}, \tag{12}$$

which satisfies the Rayleigh scaling law as the bubble radius vanishes:

$$R \propto (t_C - t)^{2/5}, t_C = \text{const}. \tag{13}$$

### C. The bounce region

As the gas inside the bubble is compressed to the hard core radius $R \to R_{min} \approx h$, the violent collapse phase is halted abruptly, and the shocklike bounce emerges. During this very short region (as shown by Fig. 2, it lasts approximately $10^{-10}$ s) the bubble wall velocity falls from supersonic speeds down to zero, releasing the energy stored in the compressed gas through emission of sound waves. The corre-sponding dynamics is governed almost exclusively by the extended gas pressure $P_{gas}$, which gives the dominant contribution to the Rayleigh-Plesset equation:

$$\rho \ddot{R} R = P_{gas} = \left(1 + \frac{R}{c}\frac{d}{dt}\right) P_g, \qquad (14)$$

where $P_g$ obeys Eq. (3). As pointed out by Löfstedt *et al.* [30], this relation yields satisfactory description of the bubble dynamics $R(t)$ for a time interval $\tau$ in the vicinity of the instant of collapse, where the value of $\tau$ is estimated as

$$\tau \approx \frac{R_{\min}}{(-dR/dt)_{\max}} \sim \frac{h}{c}. \qquad (15)$$

## III. STABILITY EQUATIONS

### A. General formulation

Let us consider an initially spherical bubble immersed in an infinite viscous liquid. In order to study the problem of shape stability, we assume a fluctuation field that perturbs the bubble-liquid interface [32]. This field of perturbations is represented by spherical Legendre polynomials as

$$\hat{R}(t,\theta,\varphi) = R(t) + \sum_{n=2}^{\infty} a_n(t) Y_n(\theta,\varphi). \qquad (16)$$

Here $R(t)$ and $\hat{R}(t,\theta,\varphi)$ are the undistorted and distorted bubble radii, respectively ($\theta$ and $\varphi$ are parameters of the spherical coordinate system whose origin is at the center of the bubble); functions $Y_n(\theta,\varphi)$ are the spherical harmonics of degree $n = (2,\ldots,\infty)$; the distortion amplitudes $a_n(t)$ are considered to be small, $|a_n(t)| \ll R(t)$. The classic example of a surface instability is the growth of perturbations on a plane interface separating a light liquid from a heavier one into which it is being uniformly accelerated; this is generally known as the Rayleigh-Taylor instability [33].

The instabilities that arise on the surface of an acoustically driven bubble are accompanied by effects related to the spherical geometry [34–38]. For the simplest case of inviscid liquids, the perturbation dynamics was derived by Plesset [34]:

$$\ddot{a}_n + \frac{3\dot{R}}{R}\dot{a}_n + (n-1)\left(-\frac{\ddot{R}}{R} + \frac{(n+1)(n+2)\sigma}{\rho R^3}\right) a_n = 0. \qquad (17)$$

The influence of liquid viscosity, being neglected in Plesset's derivation, was taken into account by Prosperetti [39]. The intrinsic difficulty of this consideration is that viscous stresses produce vorticity of the liquid in neighborhood of the bubble wall; this vorticity spreads by both convective and diffusive processes and the problem becomes strongly non-local:

$$\ddot{a}_n + \left(\frac{3\dot{R}}{R} - \frac{2(n-1)(n+1)(n+2)\mu}{\rho R^2}\right)\dot{a}_n$$
$$+ (n-1)\left(-\frac{\ddot{R}}{R} + \frac{(n+1)(n+2)(\sigma+2\mu\dot{R})}{\rho R^3}\right) a_n$$
$$- \frac{n(n+1)\dot{R}}{R^2}\int_R^{\infty}\frac{R^n}{r^n}\left(1 - \frac{R^3}{r^3}\right) T(r,t) dr$$
$$+ \frac{n(n+1)(n+2)\mu}{\rho R^2} T(R,t)$$
$$= 0. \qquad (18)$$

Here the field $T = T(r,t)$, the toroidal component of the liquid vorticity, obeys the diffusion equation

$$\frac{\partial T}{\partial t} + \dot{R}R^2\frac{\partial}{\partial r}\left(\frac{T}{r^2}\right) = \frac{\mu}{\rho}\left(\frac{\partial^2 T}{\partial r^2} - n(n+1)\frac{T}{r^2}\right) \qquad (19)$$

and the boundary condition at the bubble wall

$$T(R,t) + \frac{2}{R}\int_R^{\infty} T(r,t)\frac{R^n}{r^n} dr$$
$$= \frac{2}{n+1}\left((n+2)\dot{a}_n - (n-1)\frac{\dot{R}}{R}a_n\right). \qquad (20)$$

This exact formulation of the viscous problem is too complex for detailed analysis, although some cases of the full numerical integration of Eqs. (18)–(20) were recently reported [18,19]. In order to make the model local and more appropriate for analytical investigation, we apply several reasonable simplifications as follows. (i) The surface tension can be excluded since $\sigma \ll 2\mu\dot{R}$ for collapsing bubbles in water (this inequality, equivalent to $P_{sur} \ll P_{vis}$, is illustrated by Fig. 2 in the previous section). (ii) The integral in Eq. (18) does not contain viscous terms and, therefore, only results in tiny increments to coefficients $\dot{a}_n$ and $a_n$, in comparison with $3\dot{R}/R$ and $-\ddot{R}/R$, respectively. (iii) As an issue of simplification, one needs to approximate the viscous damping rate caused by the vorticity field. For this purpose, a boundary-layer type model was proposed by Prosperetti [40] and examined by Brener and co-workers [9,10]. According to this approach, considerable vorticity is localized within a small boundary layer of thickness $\delta$ around the bubble; then the integral in Eq. (20) is written as $2\delta T(R,t)/R$, so the vorticity at the bubble wall follows from the expression

$$T(R,t) = \frac{2}{(n+1)(1+2\delta/R)}\left((n+2)\dot{a}_n - (n-1)\frac{\dot{R}}{R}a_n\right), \qquad (21)$$

where the value of the parameter $\delta$ is given by

$$\delta = \min\left\{\sqrt{\frac{\mu}{\rho\omega}}, \frac{R}{2n}\right\}. \qquad (22)$$

Although the boundary-layer approximation is based on rather questionable assumptions (a relevant discussion is published in Refs. [16,17]), its application leads to satisfactory estimations of the viscous damping (see [18,19] and Appendix B). In this work, we consider the limiting case of a thin layer related to maximal viscous dissipation, implying the vorticity field as given in Eq. (21) with $\delta \rightarrow 0$.

By our three considerations (i)–(iii), the system of Eqs. (18)–(20) is reduced, yielding the following relation for the perturbation dynamics:

$$\ddot{a}_n + \left( \frac{3\dot{R}}{R} + \frac{2\mu(n+2)(2n+1)}{\rho R^2} \right) \dot{a}_n$$
$$+ (n-1)\left( -\frac{\ddot{R}}{R} + \frac{2\mu\dot{R}(n+2)}{\rho R^3} \right) a_n$$
$$= 0. \qquad (23)$$

### B. Perturbation dynamics of the violent collapse

In order to analyze the shape stability problem in the violent collapse region, we combine the dynamical system $a_n = a_n(R,t)$ and $t = t(R)$ [Eqs. (23) and (10)] to derive the single relation $a_n = a_n(R)$. Since Eq. (10) gives the bubble dynamics inverted [$t(R)$ instead of $R(t)$], we need to apply the formulas of conversion

$$\dot{R} = \left( \frac{dt}{dR} \right)^{-1}, \qquad (24)$$

$$\dot{a}_n = \left( \frac{dt}{dR} \right)^{-1} a_n', \qquad (25)$$

$$\ddot{R} = -\left( \frac{d^2 t}{dR^2} \right)\left( \frac{dt}{dR} \right)^{-3}, \qquad (26)$$

$$\ddot{a}_n = \left( \frac{dt}{dR} \right)^{-2} a_n'' - \left( \frac{d^2 t}{dR^2} \right)\left( \frac{dt}{dR} \right)^{-3} a_n', \qquad (27)$$

where primes denote the radial derivatives: $a_n' \equiv da_n/dR$ and $a_n'' \equiv d^2 a_n/dR^2$. The expressions for $dt/dR$ and $d^2 t/dR^2$ can be obtained from the $R$ differentiation of Eq. (10):

$$\frac{dt}{dR} = -\left( \frac{\rho}{8\mu} \right) \frac{R}{1 + \alpha \tilde{R}^{-1/2}}, \qquad (28)$$

$$\frac{d^2 t}{dR^2} = -\left( \frac{\rho}{8\mu} \right) \frac{1 + (3\alpha/2)\tilde{R}^{-1/2}}{(1 + \alpha \tilde{R}^{-1/2})^2}. \qquad (29)$$

Finally, the substitution of Eqs. (28) and (29) into Eqs. (24)–(27) and then into Eq. (23) gives the following differential relation for the perturbation dynamics (for details, see Appendix C):

$$(1 + \alpha \tilde{R}^{-1/2}) \frac{d^2 a_n}{d(\ln \tilde{R})^2} + \left( 1 + \frac{\alpha}{2}\tilde{R}^{-1/2} - \frac{(n+2)(2n+1)}{4} \right)$$
$$\times \frac{da_n}{d(\ln \tilde{R})} + (n-1)\left( \frac{3\alpha}{2}\tilde{R}^{-1/2} - \frac{(n-2)}{4} \right) a_n$$
$$= 0. \qquad (30)$$

## IV. RESULTS AND DISCUSSION

### A. Stability analysis of the violent collapse

#### 1. Theoretical investigation

In the previous section, we have derived Eq. (30), which governs the perturbation dynamics in the violent collapse region. Before we proceed to numerical simulations of this differential relation, some of its asymptotic properties can be pointed out theoretically.

(i) Let us take the inviscid limit $\alpha \tilde{R}^{-1/2} \gg 1$; in this case, Eq. (30) is reduced to

$$\frac{d^2 a_n}{d(\ln \tilde{R})^2} + \frac{1}{2}\frac{da_n}{d(\ln \tilde{R})} + \frac{3(n-1)}{2} a_n = 0. \qquad (31)$$

Then, finding a solution as $a_n = \exp(\xi \ln \tilde{R}) = \tilde{R}^\xi$ ($\xi$ is an unknown constant), one obtains

$$\xi = \pm i \frac{\sqrt{24(n-1)}}{4} - \frac{1}{4}. \qquad (32)$$

As a result, the family of solutions is represented as

$$a_n = A \tilde{R}^{-1/4} \sin[(\omega_0 \ln \tilde{R}) + \varphi_0], \qquad (33)$$

where $A$ and $\varphi_0$ are parameters determined by initial conditions, and $\omega_0 \equiv \sqrt{24(n-1) - 1/4}$. By Eq. (33), the distortions $a_n$ oscillate on the logarithmic scale of $\tilde{R}$; the amplitude of the oscillations slightly increases as the bubble collapses, obeying the relation

$$\max|a_n| \propto \tilde{R}^{-1/4}. \qquad (34)$$

(ii) In the opposite case of high viscosity $\alpha \tilde{R}^{-1/2} \ll 1$, Eq. (30) is transformed to

$$\frac{d^2 a_n}{d(\ln \tilde{R})^2} + \left( 1 - \frac{(n+2)(2n+1)}{4} \right) \frac{da_n}{d(\ln \tilde{R})}$$
$$- \frac{(n-1)(n-2)}{4} a_n = 0. \qquad (35)$$

Repeating the same procedure to find a solution, $a_n = \tilde{R}^\xi$, leads to

$$\xi = \pm \frac{1}{2}\sqrt{\left( \frac{(n+2)(2n+1)}{4} - 1 \right)^2 + (n-1)(n-2)}$$
$$+ \frac{1}{2}\left( \frac{(n+2)(2n+1)}{4} - 1 \right), \qquad (36)$$

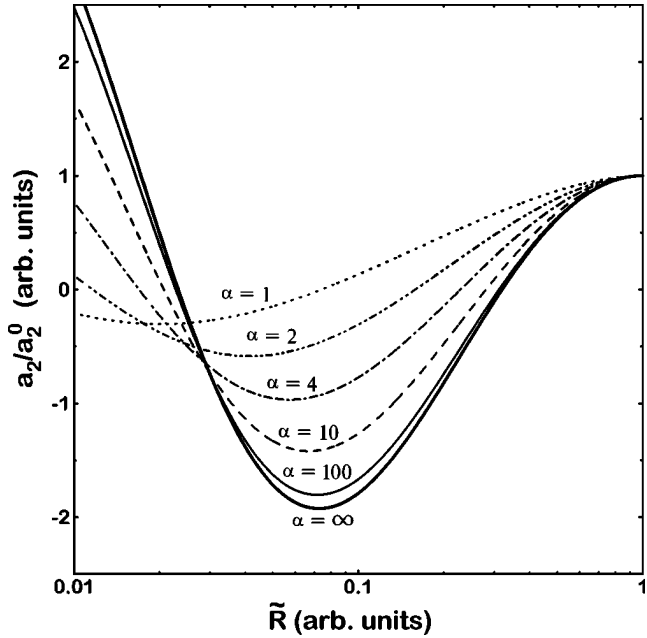

FIG. 3. Perturbation dynamics for quadrupole mode $a_2/a_2^0$ vs $\log_{10}\tilde{R}$ at various values of viscosity parameter $\alpha$=1, 2, 4, 10, 100, and $\infty$ (shown for each curve) in the violent collapse region. Initial conditions are $a_2=a_2^0$, $da_2/d\tilde{R}=0$ as $\tilde{R}=1$.

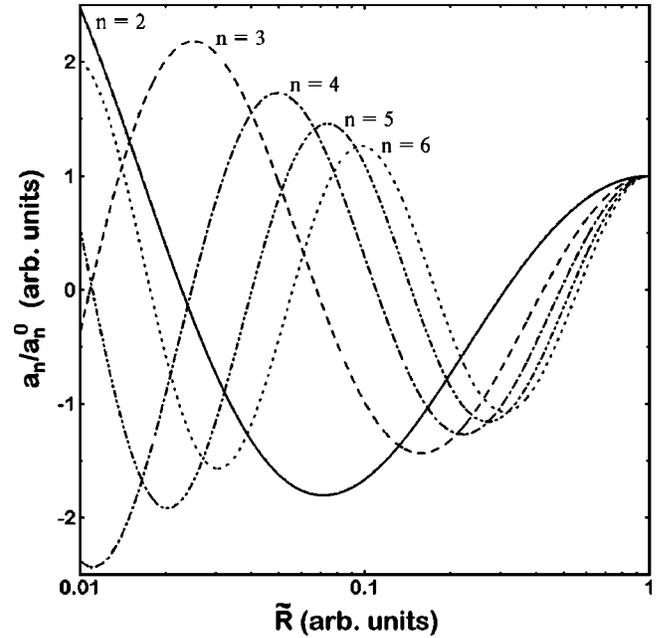

FIG. 4. Perturbation dynamics $a_n/a_n^0$ vs $\log_{10}\tilde{R}$ for modes $n$=2, . . . ,6 (shown for each curve) in the violent collapse region. Parameter of liquid viscosity $\alpha$ is fixed, $\alpha$=100. Initial conditions are $a_n=a_n^0=$const$n$, $da_n/d\tilde{R}=0$ as $\tilde{R}=1$.

which allows us to obtain the following approximation for the roots, due to the inequality $(n-1)(n-2)\ll[(n+2)(2n+1)/4-1]^2$:

$$\xi_1\approx\frac{(n+2)(2n+1)}{4}-1, \xi_2\approx-\frac{(n-1)(n-2)}{(n+2)(2n+1)-4}.$$
(37)

As a result, the family of solutions is

$$a_n=A\tilde{R}^{\xi_1}+B\tilde{R}^{\xi_2},$$
(38)

where $\xi_1$ and $\xi_2$ are defined by Eq. (37), and the parameters $A$ and $B$ are determined by initial conditions. This functional dependence means that the distortions $a_n$ obey a power-law behavior of $\tilde{R}$, giving the following scaling as the bubble radius vanishes:

$$a_n\propto\tilde{R}^{\xi_2}.$$
(39)

However, the absolute values of the negative root $\xi_2$ are small for low modes $n$ (e.g., $|\xi_2|$ rises from 0 to 0.2 as $n$ varies from 2 to 6), so the growth of perturbations by Eq. (39) is rather weak, as in the inviscid limit [Eq. (34)].

### 2. Numerical simulations

To give the overall picture for the perturbation dynamics $a_n(\tilde{R})$, results of the numerical simulations of Eq. (30) are presented in the range $0.01\leqslant\tilde{R}\leqslant1$ (in logarithmic scale of the dimensionless bubble radius $\tilde{R}$) covering all the compression stages of sonoluminescing bubbles.

In Fig. 3, we illustrate the evolution of the quadrupole mode $a_2$ for various values of the viscosity parameter $\alpha$; the initial conditions for the simulation are chosen as $a_2=a_2^0$ and

$da_2/d\tilde{R}=0$ at $\tilde{R}=1$. The calculated curves $a_2(\log_{10}\tilde{R})$ demonstrate a nonlinear oscillating behavior, where the successive increase of liquid viscosity (decrease of $\alpha$) results in a monotonic damping of the distortion amplitude. However, this viscous damping yields a substantial contribution to the perturbation dynamics only if $\alpha<10$; this inequality is not valid for sonoluminescing bubbles (where $\alpha\sim100$ due to [41]), so the influence of viscosity on the shape stability can be considered as negligible in the violent collapse region.

The perturbation dynamics $a_n(\log_{10}\tilde{R})$ for different modes $n$=2, . . . ,6 and fixed viscosity parameter $\alpha$=100 is summarized by Fig. 4; the initial conditions at $\tilde{R}=1$ are the same as in Fig. 3. As one can see by comparing Figs. 3 and 4, the high harmonics $n\geqslant3$ qualitatively resemble the dynamics of the quadrupole mode $a_2$: the obtained curves $a_n(\log_{10}\tilde{R})$ oscillate with a slight increase of amplitude as the bubble radius $\tilde{R}$ diminishes, obeying Eq. (34).

### B. The Rayleigh-Taylor instability

The stability analysis of the violent collapse region presented above has shown that the growth of the perturbation amplitude is rather weak [Eq. (34)], resulting in an insignificant contribution to the shape destabilization. This raises the following problem: does it mean that the influence of the collapse phase on subsequent development of the Rayleigh-Taylor instability is infinitesimal? In order to answer the question posed, let us study the transition from the violent collapse to the bounce in detail. When the bounce phase emerges (see Sec. II C and Fig. 2), the bubble dynamics transforms abruptly from collapsing [Eq. (10)] to shock-like [Eq. (14)]. Denoting the moment of the transformation $t = t^*$, we write the initial conditions for the posterior evolution of perturbations as

$$a_n\{t = t^*\} = a_n^*, \dot{a}_n\{t = t^*\} = \dot{a}_n^*, \tag{40}$$

where $a_n^*$ and $\dot{a}_n^*$ are the value and its time derivative of the distortion $a_n(t)$, corresponding to the end of the violent collapse region.

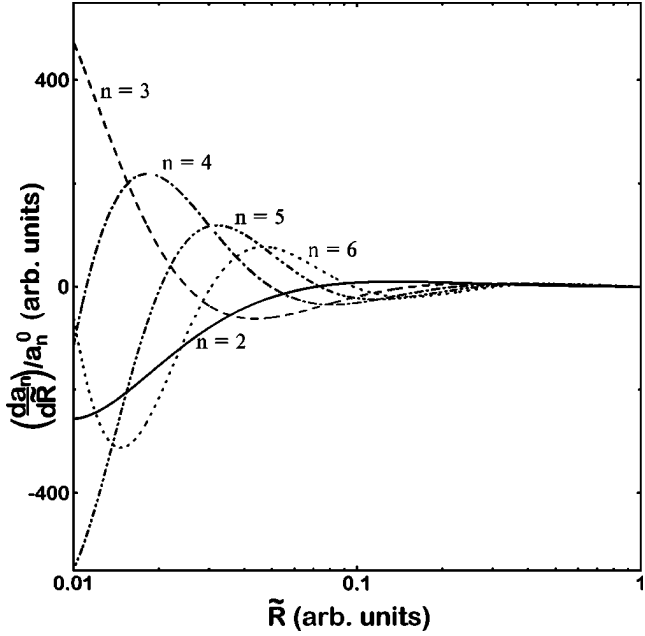

FIG. 5. Dynamics of radial derivatives $da_n/d\tilde{R}$ vs $\log_{10}\tilde{R}$ for modes $n = 2, \ldots, 6$ (shown for each curve) in the violent collapse region. Parameter of liquid viscosity $\alpha$ is fixed, $\alpha = 100$. Initial conditions are the same as in Fig. 4.

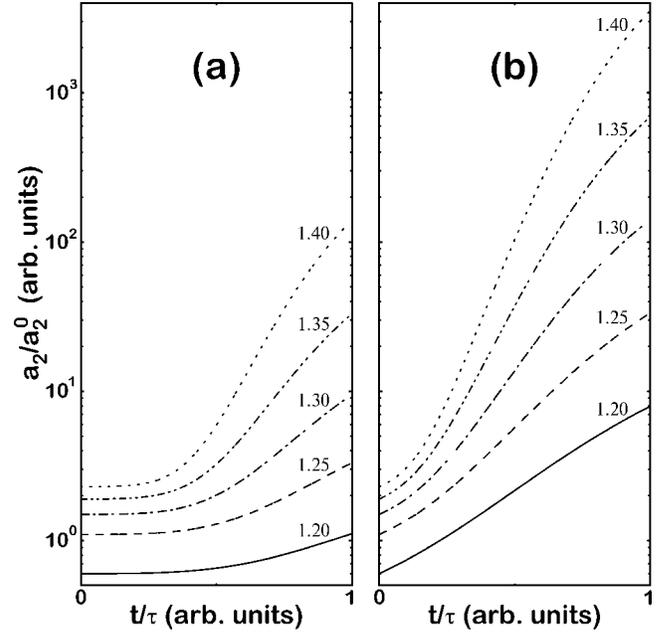

FIG. 6. Perturbation dynamics for quadrupole mode $a_2/a_2^0$ vs $t/\tau$ demonstrating intensive development of the Rayleigh-Taylor instability in the bounce region; initial derivatives $\dot{a}_2^*$ are either ignored (a) or taken into consideration (b). These are results for various values of acoustic field amplitude $P_a^0$ (atm) 1.20, 1.25, 1.30, 1.35, and 1.40 (shown for each curve). Ambient bubble radius $R_0$, ambient hydrostatic pressure $P_0$, and driving frequency $\omega$ are fixed: $R_0 = 4.5$ $\mu$m, $P_0 = 1$ atm, and $\omega = 26.5$ kHz.

From Figs. 3 and 4, the absolute values of perturbations $a_n^*$ are comparable to $a_n^0$. To find $\dot{a}_n^*$, we use the relation

$$\dot{a}_n^* = \frac{1}{R_i}\left(\frac{dR}{dt}\right)\left(\frac{da_n}{d\tilde{R}}\right) \text{ as } t \to t^*, \quad \tilde{R} \to \frac{R_{\min}}{R_i}. \tag{41}$$

The dynamics of the radial derivatives $da_n/d\tilde{R}$ vs $\log_{10}\tilde{R}$ for modes $n = 2, \ldots, 6$ is presented in Fig. 5 (obtained by the $\tilde{R}$ differentiation of Fig. 4). The curves $da_n/d\tilde{R}(\log_{10}\tilde{R})$ are characterized by strongly nonlinear oscillations with the rapid increase of amplitude as the bubble radius vanishes: the values of $|da_n/d\tilde{R}|$ achieve $500a_n^0$ for compression ratio $\tilde{R}$ of the order of $10^{-2}$. For sonoluminescing bubbles, the coefficient of $da_n/d\tilde{R}$ in Eq. (41) exceeds $10^8$ s$^{-1}$ [42], so the absolute values of $\dot{a}_n^*$ can reach $10^{11}a_n^0$ s$^{-1}$ by the final stage of the violent collapse. As a consequence, on the time scale $\tau \sim 10^{-10}$ s [Eq. (15)] the following inequality is valid:

$$|\dot{a}_n^*|\tau \gg a_n^*. \tag{42}$$

This relation means that the dominant contribution to the Rayleigh-Taylor instability stems from the time derivatives of the distortions $\dot{a}_n^*$.

The crucial role of the derivative terms $\dot{a}_n^*$ on the shape destabilization is elucidated by Fig. 6. We present evolution of the quadrupole mode $a_2(t)$ during the bounce phase $t \in [0, \tau]$ (the moment $t = 0$ corresponds to the end of the violent collapse phase). These are results for an air bubble in water: the ambient bubble radius $R_0 = 4.5$ $\mu$m, ambient hydrostatic pressure $P_0 = 1$ atm, and driving frequency $\omega = 26.5$ kHz are fixed; the amplitude of the forcing pressure $P_a^0$ is varied in the range $1.2 \leq P_a^0$ (atm) $\leq 1.4$. We assume the bubble dynamics $R(t)$ to obey Eq. (10) during the violent collapse and then Eq. (14) during the bounce phase; the parameters $a_2^*$ and $\dot{a}_2^*$ are calculated with the use of Figs. 4 and 5. Two plots are composed: (a) the initial time derivative $\dot{a}_2^*$ is ignored and (b) the term $\dot{a}_2^*$ is taken into consideration [Eq. (41)]. The figure shows the significant effect of the initial time derivatives: at the end of the time interval $t = \tau$, the values of the quadrupole mode $a_2$ with identical values of $P_a^0$ differ by more than an order of magnitude. The difference increases with the forcing pressure amplitude: for $P_a^0 = 1.4$ atm the ratio $a_2\{\dot{a}_2^* \neq 0\}/a_2\{\dot{a}_2^* = 0\}$ exceeds 30.

The results obtained allow us to discuss the development of the Rayleigh-Taylor instability quantitatively. The distortion value of the distortion $a_2^0$ relevant to the beginning of the collapse phase can be estimated as a microscopic fluctuation formed by a random displacement of magnitude $\sim 1$ nm (several diameters of the water molecule). Then the crucial perturbation of the initially spherical bubble $a_2^{<r} \sim R_{\min} \approx 0.6$ $\mu$m is related to the increase of $a_2$ by a factor of $\sim 600$ during the bounce. From Fig. 6(a), the curves $a_2(t)$ cannot achieve this threshold even as $P_a^0 = 1.4$ atm where no sonoluminescence was observed [8]. In contrast to the plot (a), the distortions $a_2(t)$ shown in Fig. 6(b) increase by three orders of magnitude, so that should result in almost full destruction of the initial bubble sphericity at $P_a^0 \geqslant 1.35$ atm, i.e., in the region where the upper threshold of the driving amplitude $P_a^0$ was experimentally reported [5].

### C. Coexistence of different instability mechanisms

Although the main reason for the shape destabilization of sonoluminescing bubbles consists in the strongest development of the Rayleigh-Taylor instability in the instant of collapse, some additional mechanisms also coexist, such as the parametric and afterbounce instabilities distinguished by Brenner and co-workers [9,10]. The first arises due to the accumulation of perturbations from sphericity over many oscillation periods, similar to Faraday waves. The second grows during the rapid afterbounces [secondary weak oscillations (Fig. 1)] that bubbles execute after the point of minimal radius. The increments of the destabilization mechanisms to the Rayleigh-Taylor instability lead to a very complex structure for the stability boundary [8–10,17–19]. Since a detailed quantitative analysis of these increments in terms of our semianalytical approach seems rather difficult, we propose a phenomenological description as follows.

From our studies, the dominant contribution to the Rayleigh-Taylor instability comes from the time derivatives of the surface distortion $\dot{a}_n^*$ at the end of the violent collapse phase. As shown in Fig. 5, the derivative terms $\dot{a}_n(\tilde{R})$ oscillate extremely nonlinear as the bubble radius $\tilde{R}$ diminishes. As a consequence, an infinitesimal shift in the initial conditions $a_n^0 \equiv a_n\{\tilde{R}=1\}$ and $\dot{a}_n^0 \equiv \dot{a}_n\{\tilde{R}=1\}$ related to the beginning of the collapse may lead to drastic change of the absolute values of $\dot{a}_n^*$ (by several orders of magnitude) or even to sign inversion. These initial conditions $a_n^0$ and $\dot{a}_n^0$ are formed during the oscillation period $T$ between the collapse moments and, therefore, are strictly influenced by the parametric and afterbounce destabilization mechanisms.

### V. SUMMARY

(i) We have shown that the general Rayleigh-Plesset equation governing the dynamics of sonoluminescing bubbles allows an analytically integrable approximation (which takes into account the liquid viscosity term) in the violent collapse region.

(ii) Based on the boundary-layer approach, we have derived a single differential relation (distortion amplitude vs bubble radius) for the perturbation dynamics during the violent collapse. Theoretical and numerical investigations reveal strongly nonlinear oscillations of the distortion amplitudes (weakly dependent on the liquid viscosity) as the bubble radius vanishes.

(iii) We have estimated the contribution of the violent collapse phase to the posterior intensive development of the Rayleigh-Taylor instability, and then have elucidated the upper threshold effect, discussing the increments from paramet-ric and afterbounce destabilization.

### ACKNOWLEDGMENTS

I would like to thank Dr. Natasha Chernova and Dr. Maxim Lobanov for stimulating discussions and helpful comments.

### APPENDIX A: DYNAMICS OF VISCOUS COLLAPSE

The proposed simplification of the general Rayleigh-Plesset dynamics [Eqs. (8) and (9)] may need additional justification, since the external pressure $P_{ext}$ exceeds (or is comparable with) the viscous term $P_{vis}$ during the violent collapse region (Fig. 2). Despite the fact that it seems desirable to include $P_{ext}$ in the approximation proposed [Eq. (8)], we have nevertheless ignored the external pressure term in our consideration for the following reasons.

(i) The function $P_{ext}(R)$ is approximately constant during the whole collapse so its increment to the bubble dynamics $R(t)$ can be compensated by an insignificant increase of the initial velocity $V_i$ in Eq. (9), in contrast to the strongly nonlinear behavior of the viscous contribution $P_{vis}$ as the bubble approaches the minimum.

(ii) The viscous pressure $P_{vis}$ is the only term in the Rayleigh-Plesset equation responsible for the dissipation of sound energy until the bounce emerges and, therefore, its consideration is preferable to that of $P_{ext}$.

(iii) The approximation introduced allows us to obtain an analytical solution of the bubble dynamics [Eq. (10)]; that yields the opportunity for subsequent detailed analysis of the shape stability problem.

### APPENDIX B: BOUNDARY-LAYER APPROXIMATION

The boundary-layer approximation was criticized by Putterman and Roberts [16] since, as they claimed, it underestimates the viscous damping rate produced by the vorticity field. They argued that the thickness $\delta$ of the boundary layer can drastically exceed the value estimated by Eq. (22), especially when the bubble reaches its minimum, yielding enhanced dissipation within the layer. Further investigations of the problem [18,19] stimulated by this criticism have revealed that the actual thickness $\delta$ is several times greater than the assumed one, but the discrepancies between the exact model and its $\delta$ approximation are nevertheless insignificant. The solution of the seeming paradox follows from simple analysis of Eq. (21): increase of the parameter $\delta$ results in a monotonic decrease of $T(R,t)$, i.e., the viscous dissipation caused by the liquid vorticity decreases as the boundary layer is enhanced. In other words, consideration of the vorticity localized cannot underestimate, but rather overestimates, the influence of liquid viscosity. The reported conditions related to slight underestimation of the viscous damping rate [18,19] are easily explained as follows. The criterion by which one can determine if the $\delta$ model overestimates (or underestimates) the viscosity effect states that the sign of $T(R,t)$ is the same as (or opposite to) the sign of $\int_R^\infty T(r,t)/r^n dr$. The function $T(r,t)$ usually demonstrates the oscillating behavior of $r$ with diminishing amplitude [18], so in most cases the signs are the same and, therefore, the boundary-layer type approximation is adequate. Some rare cases when the equality of the signs is broken, relevant to the observed discrepancies between the models, correspond to the afterbounce phase of the bubble dynamics [18]. Since in this paper we are focused on the collapse and on the bounce, the application of the $\delta$ model to our stability analysis seems rather reasonable.

### APPENDIX C: DERIVATION OF PERTURBATION DYNAMICS

The substitution of $dt/dR$ and $d^2t/dR^2$ from Eqs. (28) and (29) into Eqs. (24)–(27) yields

$$\dot{R} = -\left(\frac{8\mu}{\rho}\right)\frac{1+\alpha\widetilde{R}^{-1/2}}{R}, \tag{C1}$$

$$\dot{a}_n = -\left(\frac{8\mu}{\rho}\right)\frac{1+\alpha\widetilde{R}^{-1/2}}{R}a_n', \tag{C2}$$

$$\ddot{R} = -\left(\frac{8\mu}{\rho}\right)^2\frac{(1+\alpha\widetilde{R}^{-1/2})[1+(3\alpha/2)\widetilde{R}^{-1/2}]}{R^3}, \tag{C3}$$

$$\ddot{a}_n = \left(\frac{8\mu}{\rho}\right)^2\frac{(1+\alpha\widetilde{R}^{-1/2})^2}{R^2}a_n''$$

$$-\left(\frac{8\mu}{\rho}\right)^2\frac{(1+\alpha\widetilde{R}^{-1/2})[1+(3\alpha/2)\widetilde{R}^{-1/2}]}{R^3}a_n'. \tag{C4}$$

Then, the combination of Eqs. (C1)–(C4) with Eq. (23) gives the following:

$$R^2(1+\alpha\widetilde{R}^{-1/2})a_n'' + R\left(2 + \frac{3\alpha}{2}\widetilde{R}^{-1/2} - \frac{(n+2)(2n+1)}{4}\right)a_n'$$

$$+ (n-1)\left(\frac{3\alpha}{2}\widetilde{R}^{-1/2} - \frac{(n-2)}{4}\right)a_n = 0. \tag{C5}$$

Finally, the variable replacement $R \leftrightarrow \ln\widetilde{R}$ as

$$\frac{da_n}{dR} = \frac{1}{R}\frac{da_n}{d(\ln\widetilde{R})}, \tag{C6}$$

$$\frac{d^2a_n}{dR^2} = \frac{1}{R^2}\left(\frac{d^2a_n}{d(\ln\widetilde{R})^2} - \frac{da_n}{d(\ln\widetilde{R})}\right) \tag{C7}$$

allows us to obtain Eq. (30).

[1] D.F. Gaitan and L.A. Crum, in *Frontiers in Nonlinear Acoustics*, edited by M. Hamilton and D.T. Blackstock (Elsevier, New York, 1990), p. 459; D. F. Gaitan, Ph.D. thesis, University of Mississippi, 1990 (unpublished).

[2] D.F. Gaitan, L.A. Crum, R.A. Roy, and C.C. Church, J. Acoust. Soc. Am. **91**, 3166 (1992).

[3] B.P. Barber and S.J. Putterman, Nature (London) **352**, 318 (1991).

[4] R. Hiller, K. Weninger, S.J. Putterman, and B.P. Barber, Science **266**, 248 (1994).

[5] B.P. Barber, C.C. Wu, R. Löfstedt, P.H. Roberts, and S.J. Putterman, Phys. Rev. Lett. **72**, 1380 (1994).

[6] B.P. Barber, K. Weninger, R. Löfstedt, and S.J. Putterman, Phys. Rev. Lett. **74**, 5276 (1995).

[7] R. Löfstedt, K. Weninger, S. Putterman, and B.P. Barber, Phys. Rev. E **51**, 4400 (1995).

[8] B.P. Barber, R.A. Hiller, R. Löfstedt, S.J. Putterman, and K.R. Weninger, Phys. Rep. **281**, 65 (1997).

[9] M.P. Brenner, D. Lohse, and T.F. Dupont, Phys. Rev. Lett. **75**, 954 (1995).

[10] S. Hilgenfeldt, D. Lohse, and M.P. Brenner, Phys. Fluids **8**, 2808 (1996); **9**, 2462(E) (1997).

[11] M.P. Brenner, D. Lohse, D. Oxtoby, and T.F. Dupont, Phys. Rev. Lett. **76**, 1158 (1996).

[12] L. Kondic, C. Yuan, and C.K. Chan, Phys. Rev. E **57**, R32 (1998).

[13] S. Hilgenfeldt, D. Lohse, and W.C. Moss, Phys. Rev. Lett. **80**, 1332 (1998).

[14] R.G. Holt and D.F. Gaitan, Phys. Rev. Lett. **77**, 3791 (1996).

[15] D.F. Gaitan and R.G. Holt, Phys. Rev. E **59**, 5495 (1999).

[16] S.J. Putterman and P.H. Roberts, Phys. Rev. Lett. **80**, 3666 (1998).

[17] M.P. Brenner, T. Dupont, S. Hilgenfeldt, and D. Lohse, Phys. Rev. Lett. **80**, 3668 (1998).

[18] C.C. Wu and P.H. Roberts, Phys. Lett. A **250**, 131 (1998); Phys. Fluids **10**, 3227 (1998).

[19] Y. Hao and A. Prosperetti, Phys. Fluids **11**, 1309 (1999); A. Prosperetti and Y. Hao, Philos. Trans. R. Soc. London, Ser. A **357**, 203 (1999).

[20] V.A. Bogoyavlenskiy, Phys. Rev. E **60**, 504 (1999).

[21] Lord Rayleigh, Philos. Mag. **34**, 94 (1917).

[22] For a recent review, see S. Hilgenfeldt, M.P. Brenner, S. Grossmann, and D. Lohse, J. Fluid Mech. **365**, 171 (1998), and references therein.

[23] M.S. Plesset, J. Appl. Mech. **16**, 277 (1949).

[24] L. Trilling, J. Appl. Phys. **23**, 14 (1952).

[25] A. Prosperetti, J. Fluid Mech. **222**, 587 (1991).

[26] V. Kamath, A. Prosperetti, and F. Egolfopoulos, J. Acoust. Soc. Am. **94**, 248 (1993).

[27] C.C. Wu and P.H. Roberts, Phys. Rev. Lett. **70**, 3424 (1993); Proc. R. Soc. London, Ser. A **445**, 323 (1994).

[28] L. Kondic, J.I. Gersten, and C. Yuan, Phys. Rev. E **52**, 4976 (1995).

[29] V.Q. Vuong, A.J. Szeri, and D.A. Young, Phys. Fluids **11**, 10 (1999).

[30] R. Löfstedt, B.P. Barber, and S.J. Putterman, Phys. Fluids A **5**, 2911 (1993).

[31] R. Hiller, S.J. Putterman, and B.P. Barber, Phys. Rev. Lett. **69**, 1182 (1992); B.P. Barber and S.J. Putterman, *ibid.* **69**, 3839 (1992).

[32] H. Lamb, *Hydrodynamics*, 6th ed. (Dover Publications, New York, 1945).

[33] Lord Rayleigh, Proc. London Math. Soc. **XIV**, 170 (1883); G.I. Taylor, Proc. R. Soc. London, Ser. A **201**, 192 (1950).

[34] M.S. Plesset, J. Appl. Phys. **25**, 96 (1954).

[35] M.S. Plesset and T.P. Mitchell, Q. Appl. Math. **13**, 419 (1956).

[36] G. Birkhoff, Q. Appl. Math. **12**, 306 (1954); **13**, 451 (1956).

[37] A.I. Eller and L.A. Crum, J. Acoust. Soc. Am. **47**, 762 (1970).

[38] H.W. Strube, Acustica **25**, 289 (1971).

[39] A. Prosperetti, Q. Appl. Math. **34**, 339 (1977); Ph.D. thesis, California Institute of Technology, Pasadena, California, 1974.

[40] A. Prosperetti, Atti Accad. Naz. Lincei, Cl. Sci. Fis., Mat. Nat., Rend. **62**, 196 (1977).

[41] The beginning of the violent collapse region is defined by the condition $P_{vel} > P_{ext}$, yielding the estimation $V_i \approx 2(P_0/\rho)^{1/2} \sim 20$ m/s. Since for sonoluminescing bubbles the initial radius $R_i \sim 40$ $\mu$m, corresponding values of the liquid viscosity parameter $\alpha$ calculated from Eq. (11) are of the order of 100.

[42] Sonoluminescence is observed when the bubble wall velocity achieves supersonic speeds, $(dR/dt)_{max} = Mc$, where $c \approx 1.5 \times 10^3$ m/s is the sound speed in water and $M \sim 2-4$ is the Mach number. Since corresponding values of $R_i$ are approximately 40 $\mu$m, the expression $(dR/dt)/R_i$ is estimated as $4M \times 10^7$ s$^{-1} \sim 10^8$ s$^{-1}$.